\documentclass{elsart}
\usepackage{amssymb}
\usepackage{graphicx}

\begin{document}

\begin{frontmatter}

\title{Limit cycle bifurcations of a~reduced~Topp~system\thanksref{label1}}
\thanks[label1]{This work was supported by the London Mathematical Society
(LMS) and the Netherlands Organization for Scientific Research (NWO).
The author is very grateful to the Mathematics Institute of the University
of Warwick (UK) for hospitality during his stay in February\,--\,March 2019
and thanks Prof. Robert MacKay for kindly support and challenging discussions.}

\author{Valery A. Gaiko}

\ead{valery.gaiko@gmail.com}

\address{National~Academy~of~Sciences~of~Belarus,
United~Institute~of~Informatics~Problems,
Minsk~220012,~Belarus}

\begin{abstract}
In this paper, we carry out a global qualitative analysis of a reduced planar quartic Topp
system which models the dynamics of diabetes. In particular, studying global bifurcations,
we prove that such a system can have at most two limit cycles.
    \par
    \bigskip
\noindent \emph{Keywords}: dynamics of diabetes; Topp model; reduced planar quartic
Topp system; field rotation parameter; bifurcation; singular point; limit cycle;
Wintner--Perko termination principle
\end{abstract}

\end{frontmatter}

\section{Introduction}

In this paper, we carry out a global qualitative analysis of a reduced planar quartic
Topp system which models the dynamics of diabetes \cite{goel}, \cite{tpdmf}.

Diabetes mellitus is a disease of the glucose regulatory system characterized by fasting
or postprandial hyperglycemia. There are two major classifications of diabetes based
on the etiology of the hyperglycemia. Type~1 diabetes (also referred to as juvenile onset
or insulin-dependent diabetes) is due to an autoimmune attack on the insulin secreting $\beta$
cells. Type~2 diabetes (also referred to as adult onset or non-insulin-dependent diabetes)
is associated with a deficit in the mass of $\beta$ cells, reduced insulin secretion, and
resistance to the action of insulin; see \cite{tpdmf} and the references therein.

Blood glucose levels are regulated by two negative feedback loops. In the short term,
hyperglycemia stimulates a rapid increase in insulin release from the pancreatic $\beta$ cells.
The associated increase in blood insulin levels causes increased glucose uptake and decreased
glucose production leading to a reduction in blood glucose.
On the long term, high glucose levels lead to increase in the number of $\beta$-cells.
An increased $\beta$-cell mass
represents an increased capacity for insulin secretion which, in turn, leads to a decrease
in blood glucose. Type~2 diabetes has been associated with defects in components of both
the short-term and long-term negative feedback loops \cite{tpdmf}.

Mathematical modeling in diabetes research has focused predominately on the dynamics
of a single variable, usually blood glucose or insulin level, on a time-scale measured
in minutes \cite{tpdmf}. Generally, these models are used as tools for measuring either
rates (such as glucose production and uptake rates or insulin secretion and clearance rates)
or sensitivities (such as insulin sensitivity, glucose effectiveness, or the sensitivity of
insulin secretion rates to glucose). Two model-based studies have examined coupled glucose
and insulin dynamics \cite{tpdmf}. In each of these studies, multiple parameter changes,
representing multiple physiological defects, were required to simulate glucose and insulin
dynamics observed in humans with diabetes. In doing so, three distinct pathways were found
to the diabetic state: regulated hyperglycemia, bifurcation and dynamical hyperglycemia
\cite{tpdmf}.

In our study, we reduce the $3D$ Topp diabetes dynamics model \cite{goel}, \cite{tpdmf}
to~a~planar quartic dynamical system and study global bifurcations
of limit cycles that could occur in this system, applying the new bifurcation
methods and geometric approaches developed in \cite{bg}--\cite{gv}.
In~Section~2, we~consider the Topp model of diabetes dynamics. In~Section~3,
we~give some basic facts on singular points and limit cycles of planar dynamical systems.
In~Section~4, we carry out the global qualitative analysis of the reduced Topp system.

\section{The Topp model of diabetes dynamics}

In \cite{tpdmf}, a novel model of coupled $\beta$-cell mass, insulin, and
glucose dynamics was presented, which is used to investigate the normal behavior of the glucose regulatory
system and pathways into diabetes. The behavior of the model is consistent with the observed
behavior of the glucose regulatory system in response to changes in blood glucose levels,
insulin sensitivity, and $\beta$-cell insulin secretion rates.

In the post-absorptive state, glucose is released into the blood by the liver and kidneys,
removed from the interstitial fluid by all the cells of the body, and distributed into many
physiological compartments, e.\,g., arterial blood, venous blood, cerebral spinal fluid,
interstitial fluid \cite{tpdmf}.

Since we are primarily concerned with the evolution of fasting blood glucose levels over
a time-scale of days to years, glucose dynamics are modeled with a single-compartment
mass balance equation
    $$
    \dot{G}=a-(b+cI)G.\\[-4mm]
    \eqno(2.1)
    $$

Insulin is secreted by pancreatic $\beta$-cells, cleared by the liver, kidneys, and insulin receptors,
and distributed into several compartments, e.\,g., portal vein, peripheral blood, and interstitial
fluid. The main concern is the long-time evolution of fasting insulin levels in peripheral blood.
Since the dynamics of fasting insulin levels on this time-scale are slow, we use
a single-compartment equation given by
    $$
    \displaystyle\dot{I}=\frac{\beta G^2}{1+G^2}-\alpha I.\\[-4mm]
    \eqno(2.2)
    $$

Despite a complex distribution of pancreatic $\beta$ cells throughout the pancreas, $\beta$-cell
mass dynamics have been successfully quantified with a single-compartment model
    $$
    \dot{\beta}=(-l+mG-nG^2)\beta.\\[-4mm]
    \eqno(2.3)
    $$

Finally, the Topp model is
    $$
    \begin{array}{l}
    \dot{G}=a-(b+cI)G,
    \\[2mm]
    \displaystyle\dot{I}=\frac{\beta G^2}{1+G^2}-\alpha I,
    \\[2mm]
    \dot{\beta}=(-l+mG-nG^2)\beta
    \end{array}
    \eqno(2.4)
    $$
with parameters as in \cite{tpdmf}.

On the short timescale, $\beta$ is approximately constant and, relabelling the variables, the fast dynamics is a planar system
    $$
    \begin{array}{l}
    \dot{x}=a-(b+c\,y)x,
    \\[2mm]
    \displaystyle\dot{y}=\frac{\beta x^2}{1+x^2}-\alpha\,y
    \end{array}
    \eqno(2.5)
    $$
By rescaling time, this can be written in the form of a quartic dynamical system:
    $$
    \begin{array}{l}
\dot{x}=(1+x^2)(a-(b+c\,y)x)\equiv P,
    \\[2mm]
\dot{y}=\beta x^2-\alpha\,y(1+x^2)\equiv Q.
    \end{array}
    \eqno(2.6)
    $$
Together with (2.6), we will also consider an auxiliary system
(see \cite{BL}, \cite{Gaiko}, \cite{Perko})
    $$
\dot{x}=P-\gamma Q, \qquad \dot{y}=Q+\gamma P,
    \eqno(2.7)
    $$
applying to these systems new bifurcation methods and geometric approaches developed in
\cite{bg}--\cite{gv} and carrying out the qualitative analysis of (2.6).

\section{On singular points and limit cycles}

The study of singular points of system (2.6) will use two index theorems by H.\,Poincar\'{e};
see \cite{BL}. Let us define a singular point and its Poincar\'{e} index \cite{BL}.
    \medskip
    \par
    \noindent\textbf{Definition 3.1.}
A singular point of the dynamical system
    $$
    \dot{x}=P(x,y), \quad \dot{y}=Q(x,y),
    \eqno(3.1)
    $$
where $P(x,y)$ and $Q(x,y)$ are continuous functions (for example, polynomials),
is a point at which the right-hand sides of (3.1) simultaneously vanish.
    \medskip
    \par
    \noindent\textbf{Definition 3.2.}
Let $S$ be a simple closed curve in the phase plane not passing through a singular
point of system (3.1) and $M$ be some point on $S.$ If the point $M$ goes around
the curve $S$ once in the positive direction (counterclockwise)
then the vector coinciding with the direction of a tangent to
the trajectory passing through the point $M$ is rotated through
an angle $2\pi j$ $(j=0,\pm1,\pm2,\ldots).$ The integer $j$
is called the \emph{Poincar\'{e} index} of the closed curve $S$
relative to the vector field of system~(3.1) and has the expression
    $$
    j=\frac{1}{2\pi}\oint_S\frac{P~dQ-Q~dP}{P^2+Q^2}.
    $$
    \par

A singular point is {\em simple} if the derivative of the vector field is invertible there.
Simple singular points can be classified into nodes, foci, centers and saddles.
According to this definition, the index of a node, focus or
center is equal to $+1$, and the index of a saddle is $-1.$

A polynomial vector field on the plane can be compactified to an associated vector field on the projective plane, with a circle representing the slopes of directions to infinity (Poincar\'e compactification).  Thus we can also talk about singular points at infinity, and their indices using a local chart.
    \par
    \medskip
    \noindent\textbf{Theorem 3.1 (First Poincar\'{e} Index Theorem).}
    \emph{The indices of singular points in the plane and at infinity sum to $+1$.}
    %{If~$N,$ $N_f,$ $N_c,$ and $C$ are respectively the number
%of nodes, foci, centers, and saddles in a finite part of the phase
%plane and $N'$ and $C'$ are the number of nodes and saddles at
%infinity, then it is valid the formula}
%    $$
%    N+N_f+N_c+N'=C+C'+1.
%    $$
    \par
    \noindent\textbf{Theorem 3.2 (Second Poincar\'{e} Index Theorem).}
    \emph{If~all singular points are simple, then along an isocline
without multiple points lying in a Poincar\'{e} hemisphere which is
obtained by a stereographic projection of the phase plane
(or double cover of the projective plane), the
singular points are distributed so that a saddle is followed by
a node or a focus, or a center and vice versa. If~two~points are
separated by the equator of the Poincar\'{e} sphere, then a saddle
will be followed by a saddle again and a node or a focus, or
a center will be followed by a node or a focus, or a center.}
    \medskip
    \par
Consider polynomial system (3.1) in the vector form
    $$
    \mbox{\boldmath$\dot{x}$}=\mbox{\boldmath$f$}
    (\mbox{\boldmath$x$},\mbox{\boldmath$\mu$)},
    \eqno(3.2)
    $$
where $\mbox{\boldmath$x$}\in\textbf{R}^2;$
$\mbox{\boldmath$\mu$}\in\textbf{R}^n;$
$\mbox{\boldmath$f$}\in\textbf{R}^2$
$(\,\mbox{\boldmath$f$}$
is a polynomial vector function).
    \par
Let us state two fundamental theorems from the theory
of ana\-ly\-tic functions~\cite{Gaiko}.
    \par
    \medskip
    \noindent\textbf{Theorem 3.3 (Weierstrass Preparation Theorem).}
    \emph{Let $F(w,z)$ be an analytic function in the neighborhood of the point
$(0,0)$ satisfying the following conditions}
    $$
    F(0,0)\!=\!\frac{\partial F(0,0)}{\partial w}\!=\!\ldots\!= \frac{\partial^{k-1}F(0,0)}{\partial^{k-1}w}\!=0;\quad
    \frac{\partial^{k}F(0,0)}{\partial^{k}w}\neq0.
    $$
    \par
    \emph{Then in some neighborhood $|w|<\varepsilon,$ $|z|<\delta$ of
the point $(0,0)$, the function $F(w,z)$ can be represented as}
    $$
    F(w,z)\!=\!(w^{k}\!+A_{1}(z)w^{k-1}\!+\ldots+A_{k-1}(z)w+A_{k}(z))\Phi(w,z),
    $$
\emph{where $\Phi(w,z)$ is an analytic function not equal to zero
in the chosen neighborhood and $A_{1}(z),\ldots,A_{k}(z)$ are
analytic functions for $|z|<\delta.$}
    \par
    \medskip
From this theorem it follows that the equation $\!F(w,z)\!=\!0\!$ in a sufficiently
small neighborhood of the point $(0,0)$ is equivalent to the equation
    $$
    w^{k}+A_{1}(z)w^{k-1}+\ldots+A_{k-1}(z)w+A_{k}(z)=0,
    $$
whose left-hand side is a polynomial with respect to~$w.$ Thus,
the Weierstrass preparation theorem reduces the local study of the
general case of an implicit function $w(z),$ defined by the
equation $F(w,z)=0,$ to the case of implicit function
defined by an algebraic equation with respect to~$w.$
    \par
    \medskip
    \noindent\textbf{Theorem 3.4 (Implicit Function Theorem).}
    \emph{Let $F(w,z)$ be an analytic function in the neighborhood of
the point $(0,0)$ and $F(0,0)\!=\!0,$ $F'_{w}(0,0)\!\neq\!0.$}
    \par
    \emph{Then there exist $\delta>0$ and $\varepsilon>0$ such that
for any~$z$ satisfying the condition $|z|<\delta$ the equation
$F(w,z)=0$ has the only solution $w=f(z)$ sa\-tisfying the
condition $|f(z)|<\varepsilon.$ The func\-tion $f(z)$ is
expanded into the series on positive integer powers of~$z$ which
converges for $|z|\!<\!\delta,$ i.\,e., it is a single-valued analytic
function of~$z$ which vanishes at $z=0.$}
    \medskip
    \par
Let us recall some basic facts concerning limit cycles of (3.2).
Assume that system (3.2) has a limit cycle
    $$
    L_0:\mbox{\boldmath$x$}=\mbox{\boldmath$\varphi$}_0(t)
    $$
of minimal period $T_0$ at some parameter value
$\mbox{\boldmath$\mu$}\!=\!\mbox{\boldmath$\mu$}_0\!\in\textbf{R}^n;$
see~Fig.~1 \cite{Gaiko}.
    \par
    \begin{figure}[htb]
\begin{center}
    \includegraphics[width=138.5mm]{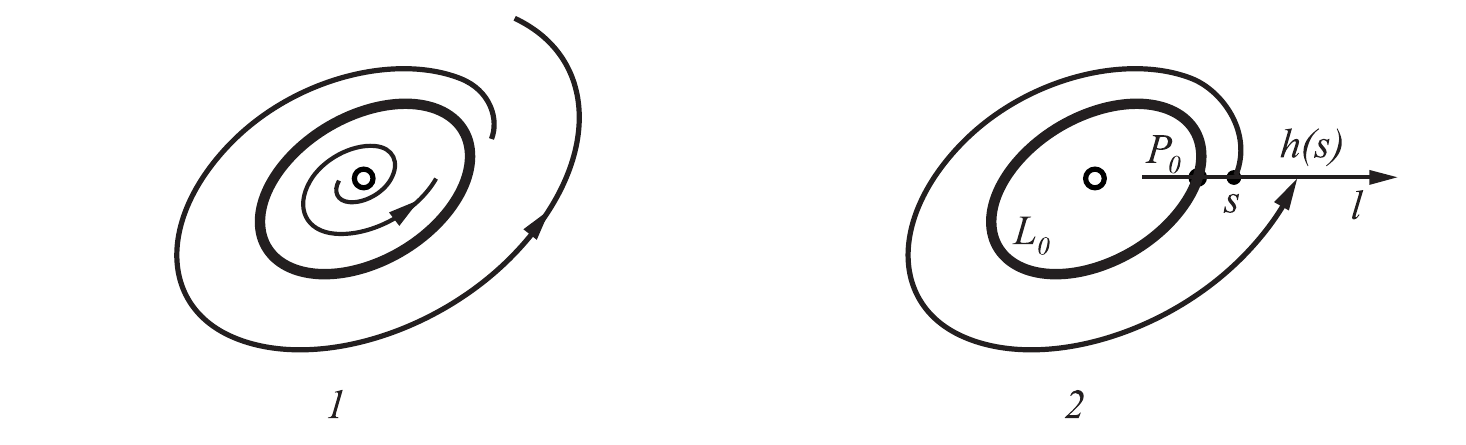}
    \vspace{-4mm}
    \par
    {\small {\bf Figure~1.} The Poincar\'{e} return map
    in the neighborhood of a multiple limit cycle.}
\end{center}
    \end{figure}
    \medskip
    \par
Let~$l$ be the straight line normal to $L_0$ at the
point $\mbox{\boldmath$p$}_0=\mbox{\boldmath$\varphi$}_0(0)$ and
$s$ be the coordinate along $l$ with $s$ positive exterior to
$L_0.$ It then follows from the implicit function theorem that
there is a $\delta>0$ such that the Poincar\'{e} map
$h(s,\mbox{\boldmath$\mu$})$ is defined and analytic for $|s|<\delta$
and $\|\mbox{\boldmath$\mu$}-\mbox{\boldmath$\mu$}_0\|<\delta.$
The displacement function for system (3.2) along
the normal line~$l$ to~$L_0$ is defined as the function
    $$
    d(s,\mbox{\boldmath$\mu$})=h(s,\mbox{\boldmath$\mu$})-s.
    $$
    \par
We denote derivatives of $d$ with respect to $s$ or components of $\mu$ by subscripts, and the $m^{th}$ derivative of $d$ with respect to $s$ by $d_s^{(m)}$.
In terms of the displacement function, a multiple limit cycle
can be defined as follows \cite{Gaiko}.
    \par
    \medskip
    \noindent\textbf{Definition 3.3.}
A limit cycle $L_0$ of (3.2) is a \emph{multiple limit cycle} iff
$d(0,\mbox{\boldmath$\mu$}_0)\!=\!d_s(0,\mbox{\boldmath$\mu$}_0)\!=\!0$.
It is a \emph{simple limit cycle} (or hyperbolic limit cycle)
if it is not a multiple limit cycle; furthermore, $L_0$ is a limit
cycle of multiplicity~$m$ iff
    $$
    d(0,\mbox{\boldmath$\mu$}_0)=d_s(0,\mbox{\boldmath$\mu$}_0)=\ldots
    =d_s^{(m-1)}(0,\mbox{\boldmath$\mu$}_0)=0,
    $$
    $$
    d_s^{(m)}(0,\mbox{\boldmath$\mu$}_0)\neq 0.
    $$
    \par
Note that the multiplicity of $L_0$ is independent of the point
$\mbox{\boldmath$p$}_0\in L_0$ through which we take the normal
line~$l.$
    \par
Let us write down also the following formulae which have already
become classical ones and determine the derivatives of the
displacement function in terms of integrals of the vector
field~$\mbox{\boldmath$f$}$ along the periodic orbit
$\mbox{\boldmath$\varphi$}_0(t)$ \cite{Gaiko}:
    \vspace{-2mm}
    $$
    d_s(0,\mbox{\boldmath$\mu$}_0)\;=\;\displaystyle\exp\int_{0}^{T_0}\!
    \mbox{\boldmath$\nabla$}\cdot\mbox{\boldmath$f$}
    (\mbox{\boldmath$\varphi$}_0(t),\mbox{\boldmath$\mu$}_0)\:\textrm{d}t-1
    $$
and
    $$
    d_{\mu_j}(0,\mbox{\boldmath$\mu$}_0)=
    \frac{-\omega\,_0}{\|\mbox{\boldmath$f$}
    (\mbox{\boldmath$\varphi$}_0(0),
    \mbox{\boldmath$\mu$}_0)\|}\,\times$$
    $$
    \displaystyle\int_{0}^{T_0}\!\!
    \exp\left(-\!\int_{0}^{t}\!\mbox{\boldmath$\nabla$}\cdot
    \mbox{\boldmath$f$}(\mbox{\boldmath$\varphi$}_0(\tau),
    \mbox{\boldmath$\mu$}_0)\,\textrm{d}\tau\right)\times
    \mbox{\boldmath$f$}\wedge\mbox{\boldmath$f$}_{\mu_j}
    (\mbox{\boldmath$\varphi$}_0(t),\mbox{\boldmath$\mu$}_0)\:\textrm{d}t
    \\[2mm]
    $$
for $j=1,\ldots,n,$ where $\omega_0=\pm1$ according to whether $L_0$
is positively or negatively oriented, respectively, and where the wedge
product of two vectors $\mbox{\boldmath$x$}=(x_1,x_2)$ and
$\mbox{\boldmath$y$}=(y_1,y_2)$ in $\textbf{R}^2$ is defined as
    $$
    \mbox{\boldmath$x$}\wedge\mbox{\boldmath$y$}=x_1\,y_2-x_2\,y_1.
    \vspace{-1mm}
    $$
    \par
Similar formulae for $d_{ss}(0,\mbox{\boldmath$\mu$}_0)$ and
$d_{s{\mu_j}}(0,\mbox{\boldmath$\mu$}_0)$ can be derived in terms
of integrals of the vector field $\mbox{\boldmath$f$}$ and its first
and second partial derivatives along $\mbox{\boldmath$\varphi$}_0(t).$
    \par
Now we can formulate the Wintner--Perko termination principle \cite{Perko}
for polynomial system (3.2).
    \par
    \medskip
    \noindent\textbf{Theorem 3.5 (Wintner--Perko Termination Principle).}
    \emph{Any one-para\-me\-ter fa\-mi\-ly of multip\-li\-city-$m$
limit cycles of relatively prime polynomial system $(3.2)$ can be
extended in a unique way to a maximal one-parameter family of
multiplicity-$m$ limit cycles of $(3.2)$ which is either open
or cyclic.}
    \par
\emph{If it is open, then it terminates either as the parameter or
the limit cycles become unbounded; or, the family terminates
either at a singular point of $(3.2),$ which is typically
a fine focus of multiplicity~$m,$ or on a (compound)
separatrix cycle of $(3.2)$ which is also typically of
multiplicity~$m.$}
    \medskip
    \par
The proof of this principle for general polynomial system (3.2) with
a vector parameter $\mbox{\boldmath$\mu$}\in\textbf{R}^n$ parallels
the proof of the pla\-nar termination principle for the system
    $$
    \vspace{1mm}
    \dot{x}=P(x,y,\lambda),
        \quad
    \dot{y}=Q(x,y,\lambda)
    \eqno(3.3)
    $$
with a single parameter $\lambda\in\textbf{R}$ (see \cite{Gaiko},
\cite{Perko}), since there is no loss of generality in assuming
that system (3.2) is parameterized by a single parameter $\lambda;$
i.\,e., we can assume that there exists an analytic mapping
$\mbox{\boldmath$\mu$}(\lambda)$ of $\textbf{R}$ into $\textbf{R}^n$
such that (3.2) can be written as (3.3) and then we can repeat everything
that had been done for system (3.3) in \cite{Perko}.
In particular, $\lambda$ is said to be a {\em field-rotation parameter}
if it rotates the vectors of the field in one direction \cite{BL}, \cite{Gaiko}, \cite{Perko}, e.g.~$\gamma$ in (2.7).
If $\lambda$ is a field rotation parameter of (3.3), the following theorem of Perko
on monotonic families of limit cycles is valid; see \cite{Perko}.
    \par
    \medskip
    \noindent\textbf{Theorem 3.6.}
    \emph{If $L_{0}$ is a nonsingular multiple limit cycle of $(3.3)$
for $\lambda=\lambda_{0},$ then $L_{0}$ belongs to a one-parameter family
of limit cycles of $(3.3);$ furthermore$:$}
    \par
1)~\emph{if the multiplicity of $L_{0}$ is odd, then the family
either expands or contracts mo\-no\-to\-ni\-cal\-ly as $\lambda$
increases through $\lambda_{0};$}
    \par
2)~\emph{if the multiplicity of $L_{0}$ is even, then $L_{0}$
bi\-fur\-cates into a stable and an unstable limit cycle as
$\lambda$ varies from $\lambda_{0}$ in one sense and $L_{0}$
dis\-ap\-pears as $\lambda$ varies from $\lambda_{0}$ in the
opposite sense; i.\,e., there is a fold bifurcation at
$\lambda_{0}.$}

\section{Global bifurcation analysis}

Consider system (2.6). Its finite singularities are determined by the algebraic system
   $$
    \begin{array}{l}
(1+x^2)(a-(b+c\,y)x)=0,
    \\[2mm]
\beta x^2-\alpha\,y(1+x^2)=0
    \end{array}
    \eqno(4.1)
    $$
which can give us at most three singular points in the first quadrant: a saddle $S$ and
two antisaddles (non-saddles)~--- $A_{1}$ and $A_{2}$ --- according to the second Poincar\'{e} index theorem
(Theorem 3.2). Suppose that with respect to the $x$-axis they have the following sequence:
$A_{1},$ $S,$ $A_{2}.$ System (2.6) can also have one singular point (an antisaddle)
or two singular points (an antisaddle and a saddle-node) in the first quadrant.
    \par
To study singular points of (2.6) at infinity, consider the corresponding diffe\-rential equation
    $$
\frac{dy}{dx}=\frac{\beta x^2-\alpha\,y(1+x^2)}
{(1+x^2)(a-(b+c\,y)x)}.
    \eqno(4.2)
    $$
  \par
Dividing the numerator and denominator of the right-hand side of (4.2) by $x^{4}$ $(x\neq0)$
and denoting $y/x$ by $u$ (as well as $dy/dx),$ we will get the equation
    $$
u^2=0, \quad \mbox{where} \quad u=y/x,
    \eqno(4.3)
    $$
for all infinite singularities of (4.2) except when $x=0$ (the ``ends'' of the $y$-axis);
see \cite{BL}, \cite{Gaiko}. For this special case we can divide the numerator and denominator
of the right-hand side of (4.2) by $y^{4}$ $(y\neq0)$ denoting $x/y$ by $v$ (as well as $dx/dy)$ and consider the equation
    $$
v^2=0, \quad \mbox{where} \quad v=x/y.
    \eqno(4.4)
    $$
According to the Poincar\'{e} index theorems (Theorem~3.1 and Theorem~3.2), the equations (4.3)
and (4.4) give us two double singular points (saddle-nodes) at infinity for (4.2): on the ``ends''
of the $x$ and $y$ axes.
    \par
Using the obtained information on singular points and applying geometric approaches developed in
\cite{bg}--\cite{gv}, we can study now the limit cycle bifurcations of system (2.6).
    \par
Applying the definition of a field rotation parameter \cite{BL}, \cite{Gaiko}, \cite{Perko},
to system (2.6), let us calculate the corresponding determinants for the parameters $a,$~$b,$~$c,$
$\alpha,$ and $\beta,$ respectively:
    $$
\Delta_{a}=PQ'_{a}-QP'_{a}=-(1+x^2)(\beta x^2-\alpha\,y(1+x^2)),\\
    \eqno(4.5)
    $$
    $$
\Delta_{b}=PQ'_{b}-QP'_{b}=x(1+x^2)(\beta x^2-\alpha\,y(1+x^2)),\\[2mm]
    \eqno(4.6)
    $$
    $$
\Delta_{c}=PQ'_{c}-QP'_{c}=xy(1+x^2)(\beta x^2-\alpha\,y(1+x^2)),\\[2mm]
    \eqno(4.7)
    $$
    $$
\Delta_{\alpha}=PQ'_{\alpha}-QP'_{\alpha}=-y(1+x^2)^2(a-(b+c\,y)x),\\[2mm]
    \eqno(4.8)
    $$
    $$
\Delta_{\beta}=PQ'_{\beta}-QP'_{\beta}=x^2(1+x^2)(a-(b+c\,y)x).\\[4mm]
    \eqno(4.9)
    $$
It follows from (4.5)--(4.7) that in the first quadrant the signs of $\Delta_{a},$ $\Delta_{b},$ $\Delta_{c}$ depend on the sign of $\beta x^2-\alpha\,y(1+x^2)$ and from (4.8) and (4.9)
that the signs of $\Delta_{\alpha}$ and $\Delta_{\beta}$ depend on the sign of $a-(b+c\,y)x$
on increasing (or~decreasing) the parameters $a,$ $b,$ $c,$ $\alpha,$ and $\beta,$ respectively.
    \par
Therefore, to study limit cycle bifurcations of system (2.6), it makes sense together
with (2.6) to consider also the auxiliary system (2.7) with field-rotation
parameter~$\gamma\!:$
    $$
\Delta_{\gamma}=P^{2}+Q^{2}\geq0.
    \eqno(4.10)
    $$
    \par
Using system (2.7) and applying Perko's results, we will prove the following theorem.
	\par
    \medskip
\noindent \textbf{Theorem 4.1.}
\emph{The reduced Topp system $(2.6)$ can have at most two limit cycles.}
	  \par
    \medskip
\noindent\textbf{Proof.} In \cite{bnrs}, \cite{bg}, \cite{lx}, \cite{zcw}, where a similar quartic system was studied, it was proved that the cyclicity of singular points in such a system is equal
to two and that the system can have at least two limit cycles; see also \cite{gai5}, \cite{gv},
\cite{gmrf}, \cite{lcr} with similar results.
	  \par
Consider systems (2.6)--(2.7) supposing that the cyclicity of singular points in these systems
is equal to two and that the systems can have at least two limit cycles. Let us prove now that
these systems have at most two limit cycles. The proof is carried out by contradiction applying
Catastrophe Theory; see~\cite{Gaiko},~\cite{Perko}.
    \par
We will study more general system (2.7) with three parameters: $\alpha,$ $\beta,$ and $\gamma$
(the parameters $a,$ $b,$ and $c$ can be fixed, since they do not generate limit cycles).
Suppose that (2.7) has three limit cycles surrounding the singular point $A_{1},$ in the first
quadrant. Then we get into some domain of the parameters $\alpha,$ $\beta,$ and $\gamma$
being restricted by definite con\-di\-tions on three other parameters, $a,$ $b,$ and $c$.
This domain is bounded by two fold bifurcation surfaces forming a cusp bifurcation surface
of multiplicity-three limit cycles in the space of the pa\-ra\-me\-ters $\alpha,$ $\beta,$
and $\gamma;$ see~Fig.~2 \cite{Gaiko}.
    \par
\begin{figure}[htb]
\begin{center}
\includegraphics[width=100mm]{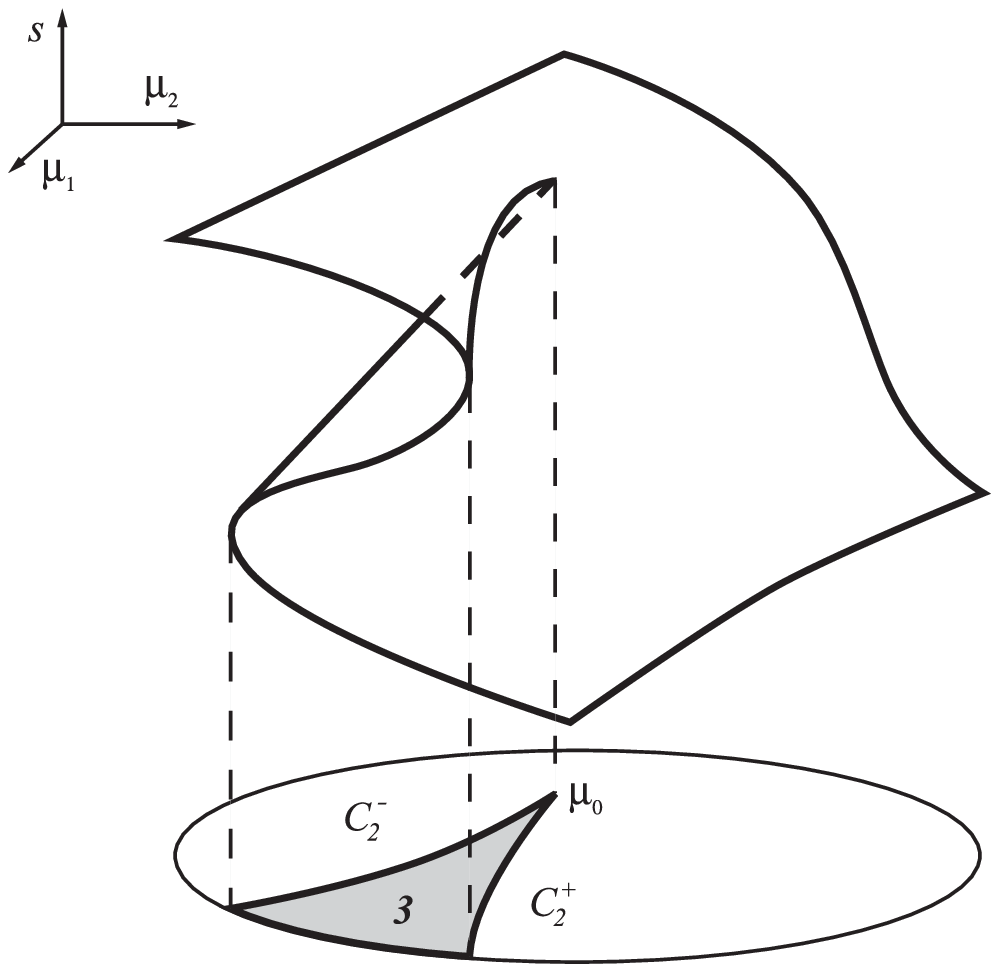}
    \vspace{4mm}
    \par
{\small {\bf Figure~2.} The cusp bifurcation surface.}\\[2mm]
\end{center}
\end{figure}
    \par
The cor\-res\-pon\-ding maximal one-parameter family of multiplicity-three limit cycles cannot
be cyclic, otherwise there will be at least one point cor\-res\-pon\-ding to the limit cycle of
multi\-pli\-ci\-ty four (or even higher) in the parameter space; see~Fig.~3 \cite{Gaiko}.
    \par
\begin{figure}[htb]
\begin{center}
\includegraphics[width=100mm]{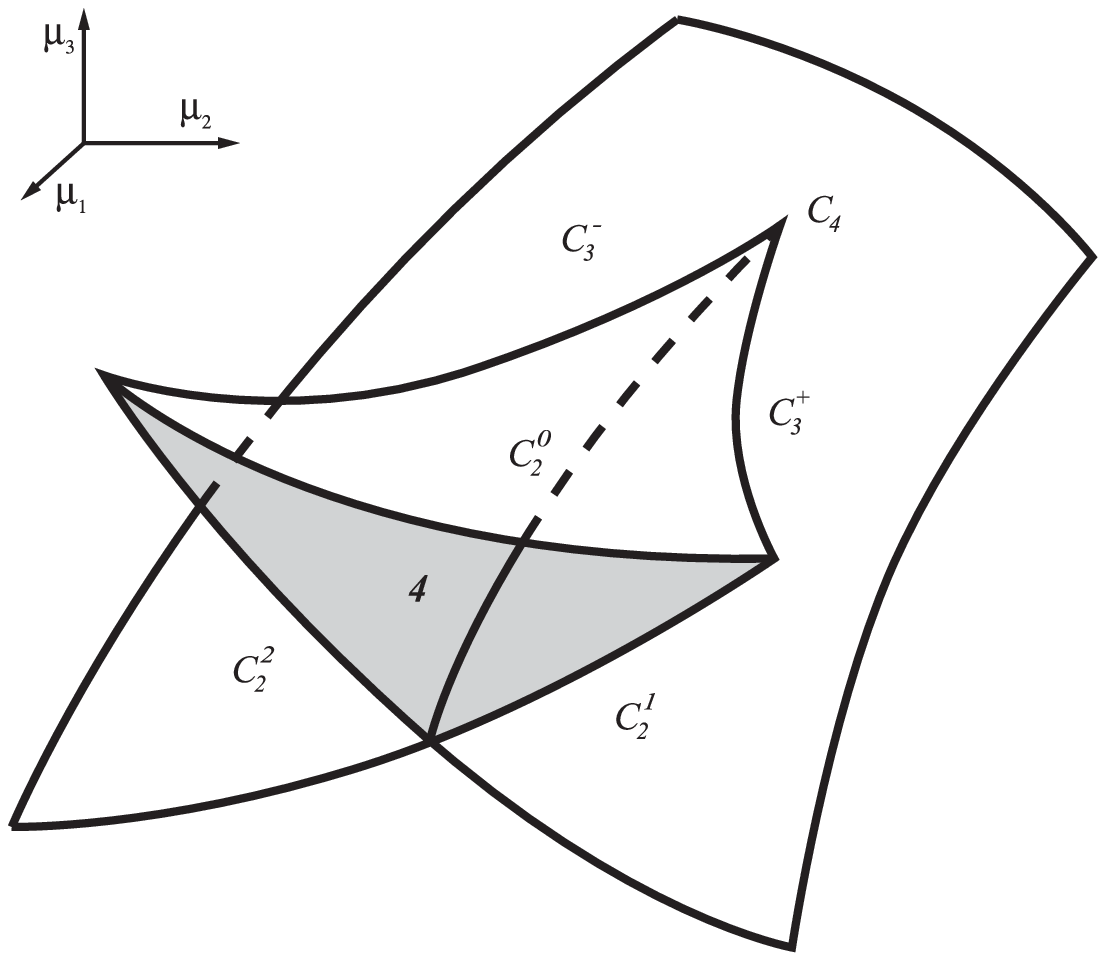}
    \par
{\small {\bf Figure~3.} The swallow-tail bifurcation surface.}
\end{center}
\end{figure}
    \par
Extending the bifurcation curve of multi\-pli\-ci\-ty-four limit cycles
through this point and parameterizing the corresponding maximal one-parameter
family of multi\-pli\-ci\-ty-four limit cycles by the field rotation para\-me\-ter,
$\gamma,$ according to Theorem~3.6, we will obtain two monotonic curves of
multi\-pli\-ci\-ty-three and one, respectively, which, by the Wintner--Perko termination
principle (Theorem~3.5), terminate either at the point $A_{1}$ or on a separatrix cycle
surrounding this point. Since on our assumption the cyclicity of the singular point is
equal to two, we have obtained a contradiction with the termination principle stating
that the multiplicity of limit cycles cannot be higher than the multi\-pli\-ci\-ty
(cyclicity) of the singular point in which they terminate; see~Fig.~4 \cite{Gaiko}.
    \par
    \begin{figure}[htb]
\begin{center}
\includegraphics[width=100mm]{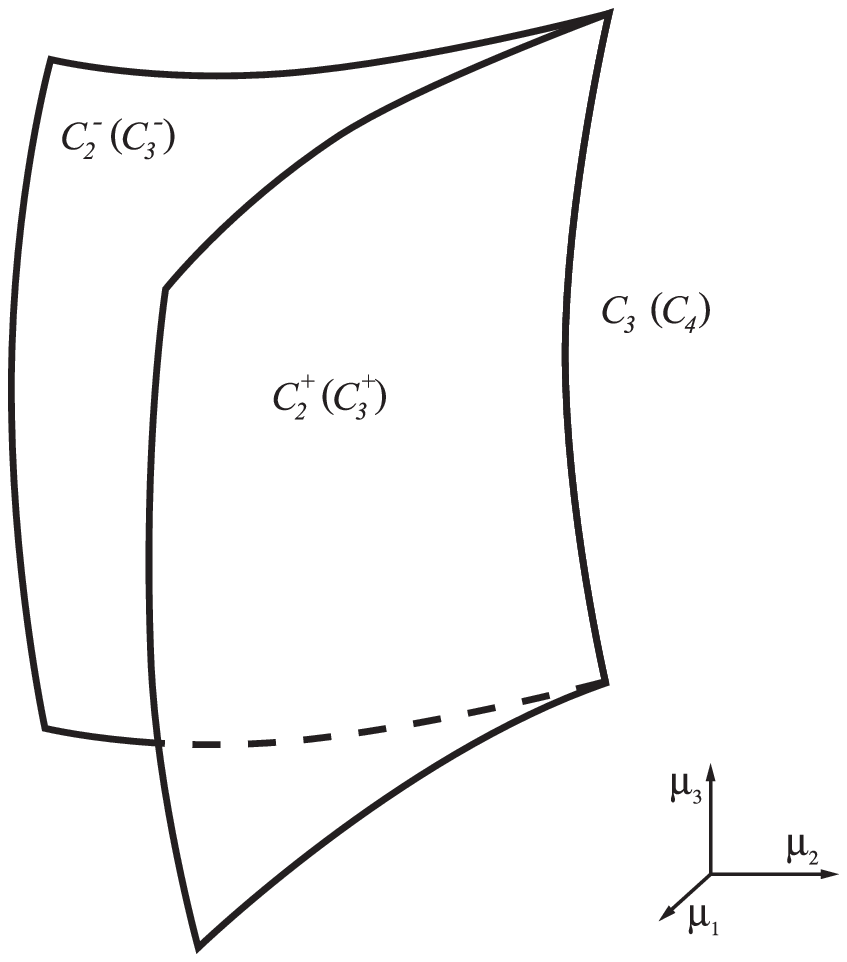}
    \par
{\small {\bf Figure~4.} The bifurcation curve (one-parameter family)
of multiple limit cycles.}
\end{center}
\end{figure}
    \par
If the maximal one-parameter family of multiplicity-three limit cycles is not cyclic,
using the same principle (Theorem~3.5), this again contradicts the cyclicity of $A_{1}$
not admitting the multiplicity of limit cycles to be higher than two. This contradiction
completes the proof in the case of one singular point in the first quadrant.
    \par
Suppose that system (2.7) with three finite singularities, $A_{1},$ $S,$ and $A_{2},$ has
two small limit cycles around, for example, the point $A_{1}$ (the case when limit cycles
surround the point $A_{2}$ is considered in a similar way). Then we get into some domain
in the space of the parameters $\alpha,$ $\beta,$ and $\gamma$ which is bounded by a fold
bifurcation surface of multiplicity-two limit cycles; see~Fig.~5 \cite{Gaiko}.
    \par
\begin{figure}[htb]
\begin{center}
\includegraphics[width=100mm]{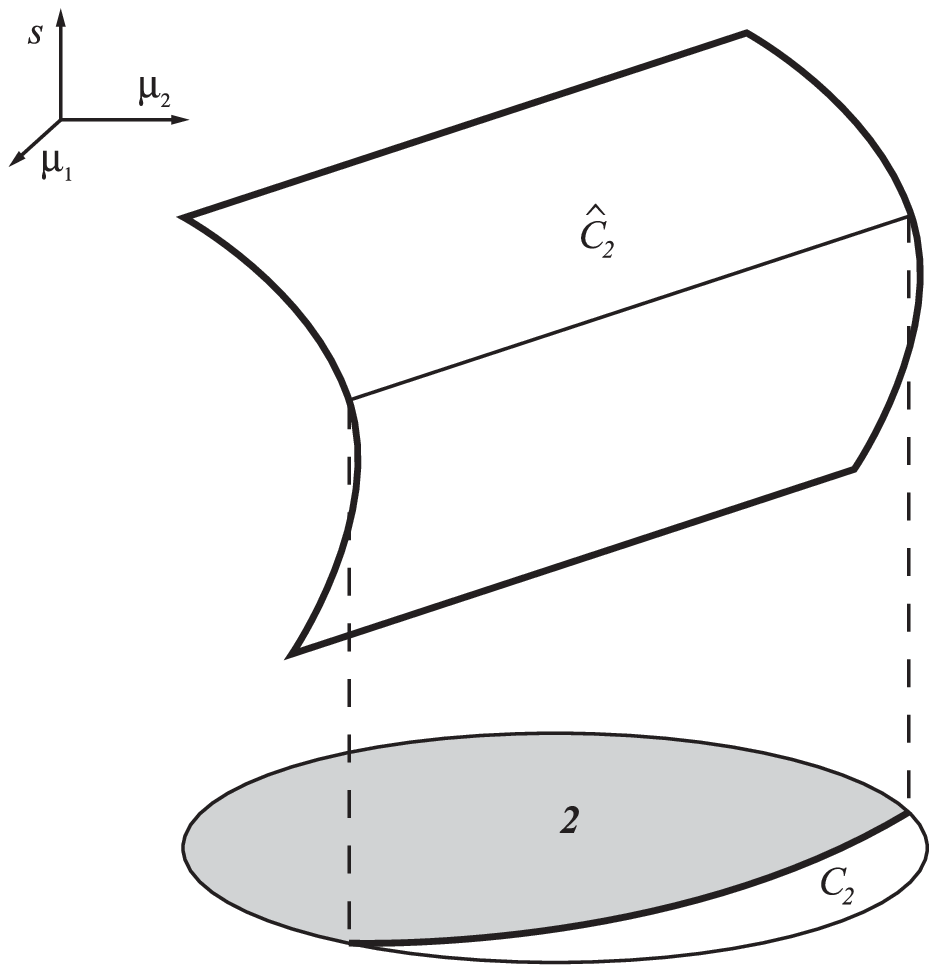}
    \vspace{4mm}
    \par
{\small {\bf Figure~5.} The fold bifurcation surface.}\\[2mm]
\end{center}
\end{figure}
    \par
The cor\-res\-pon\-ding maximal one-parameter family of multiplicity-two limit cycles cannot
be cyclic, otherwise there will be at least one point cor\-res\-pon\-ding to the limit cycle
of multi\-pli\-ci\-ty three (or even higher) in the parameter space. Extending the bifurcation
curve of multi\-pli\-ci\-ty-three limit cycles through this point and parameterizing the corresponding maximal one-parameter family of multi\-pli\-ci\-ty-three limit cycles by the
field rotation para\-me\-ter, $\gamma,$ according to Theorem~3.6, we will obtain a monotonic
curve which, by the Wintner--Perko termination principle (Theorem~3.5), terminates either at
the point $A_{1}$ or on some separatrix cycle surrounding this point. Since we know at least
the cyclicity of the singular point which on our assumption is equal to one in this case, we
have obtained a contradiction with the termination principle.
    \par
If the maximal one-parameter family of multiplicity-two limit cycles is not cyclic, using
the same principle (Theorem~3.5), this again contradicts the cyclicity of $A_{1}$ not admitting
the multiplicity of limit cycles higher than one. Moreover, it also follows from the termination
principle that either an ordinary (small) separatrix loop or a big loop, or an eight-loop cannot
have the multiplicity (cyclicity) higher than one in this case. Therefore, according to the same
principle, there are no more than one limit cycle in the exterior domain surrounding all three
finite singularities, $A_{1},$ $S,$ and $A_{2}.$
    \par
Thus, taking into account all other possibilities for limit cycle bifurcations (see \cite{bnrs}, \cite{bg}, \cite{lx}, \cite{zcw}), we conclude that system~(2.7) (and (2.6) as well) cannot have
either a multiplicity-three limit cycle or more than two limit cycles in any configuration.
The theorem is proved.
\qquad $\Box$

\end{document}